**Title**

Evaluation of dental implant stability in bone phantoms: comparison between a quantitative ultrasound technique and resonance frequency analysis

**Running Head (abbreviated title)**

RFA and ultrasound methods to evaluate implant stability

**Authors**

Romain VAYRON, PhD, Postdoctoral researcher ; CNRS, Laboratoire Modélisation et Simulation MultiEchelle, MSME UMR CNRS 8208, 61, avenue du Général de Gaulle, 94010 Créteil, Cedex, France

Vu Hieu NGUYEN, PhD, Associate Professor ; CNRS, Laboratoire Modélisation et Simulation MultiEchelle, MSME UMR CNRS 8208, 61, avenue du Général de Gaulle, 94010 Créteil, Cedex, France

Benoît LECUELLE, Bac+2, Engineer Assist ; Centre de Recherche BioMédicale, Ecole Nationale Vétérinaire d'Alfort, 7 Avenue du Général de Gaulle, 94700 Maisons-Alfort France

Guillaume HAIAT, PhD, Research Director ; CNRS, Laboratoire Modélisation et Simulation MultiEchelle, MSME UMR CNRS 8208, 61, avenue du Général de Gaulle, 94010 Créteil, Cedex, France

**Corresponding Author**

Guillaume HAÏAT
Laboratoire de Modélisation et de Simulation MultiEchelle, UMR CNRS 8208,
61 avenue du Général de Gaulle,
94010 Créteil, France
tel : (33) 1 45 17 14 41
fax : (33) 1 45 17 14 33
e-mail : guillaume.haiat@univ-paris-est.fr

# Abstract

*Background*

Resonance frequency analyses and quantitative ultrasound methods have been suggested to assess dental implant primary stability.

*Purpose*

The purpose of this study was to compare the results obtained using these two techniques applied to the same dental implants inserted in various bone phantoms.

*Materials and Methods*

Different values of trabecular bone density and cortical thickness were considered to assess the effect of bone quality on the respective indicators (*UI and ISQ*). The effect of the implant insertion depth and of the final drill diameter was also investigated.

*Results*

*ISQ* values increase and *UI* values decrease as a function of trabecular density, cortical thickness and the screwing of the implant. When the implant diameter varies, the *UI* values are significantly different for all final drill diameters (except for two), while the *ISQ* values are similar for all final drill diameters lower than 3.2 mm and higher than 3.3 mm. The error on the estimation of parameters with the QUS device is between 4 and 8 times lower compared to that made with the RFA technique.

*Conclusions*

The results show that ultrasound technique provides a better estimation of different parameters related to the implant stability compared to the RFA technique.

# Introduction

Dental implant stability, which is determinant for the surgical success [1], is determined by the quantity and biomechanical quality of bone tissue around the implant [2]. Two kinds of implant stability may be distinguished. The primary stability occurs at the moment of implant surgical insertion within bone tissue. Dental implant primary stability should be sufficiently high in order to avoid micromotion at the bone-implant interface after surgery, but should not be too high to avoid bone necrosis due to overloading of bone tissue [3]. Secondary stability is obtained through osseointegration phenomena, a complex phenomenon of a multi-time and multiscale nature [3], which strongly depends on the implant primary stability.

Dental implant stability remains difficult to be assessed clinically because it depends on the implant properties (geometry, surface properties...), on the patient bone quality, as well as on the surgical protocol. In particular, there is a lack of standardization of the surgical procedures used in dental implantology, for example in the choice of the duration between implant insertion and loading, which may vary from 0 up to 6 months [4]. A compromise should be found in a patient specific manner between i) an early (or even immediate [5]) implant loading in order to stimulate osseointegration phenomena and ii) a late implant loading in order to avoid degradation of the consolidating bone-implant interface in early postsurgical stages [6]. Meanwhile, shortening the time of implant loading has become a challenge in recent implant developments to i) minimize the time of social disfigurement and ii) avoid gum loss. As a consequence, accurate measurements of implant biomechanical stability are of interest since they could be used to improve the surgical strategy by adapting the choice of the healing period in a patient-specific manner.

Assessing the implant stability is a difficult multiscale problem because of the complex heterogeneous nature of periprosthetic bone tissue and to remodeling phenomena [7, 8]. Different approaches have been used to assess the implant stability *in vivo*. So far, most surgeons still rely on their proprioception because it remains difficult to monitor bone healing *in vivo* [6]. Accurate quantitative methods capable of assessing implant stability are required to guide the surgeons and eventually reduce the risk of implant failure.

Magnetic resonance imaging [9] and X-ray based [10] techniques remain of limited interest to measure implant stability because of diffraction phenomena occurring at the bone-implant interface due to the presence of metal. Therefore, biomechanical methods have been developed, their main advantage consisting in the absence of ionizing radiation, inexpensiveness, portability and noninvasiveness. The measurement of the insertion torque to assess dental implant primary stability has been evoked but such approach remains limited [11] because the result is not only related to the properties of the bone-implant interface and because it cannot be used for secondary stability assessment. The Periotest (Bensheim, Germany) is a percussion test methods [12, 13]. Its sensitivity to striking height and handpiece angulation complicates the clinical examination [14] and limits the reproducibility of the measurements. The most commonly used biomechanical technique is the resonance frequency analysis (RFA) [15], which consists in measuring [16] the first bending resonance frequency of a small rod attached to the implant. The RFA technique allows to assess the implant anchorage depth into bone [17], marginal bone level [18] and the stiffness of the bone-implant structure [19, 20]. However, RFA cannot be used to identify directly the bone-implant interface characteristics [21]. No correlation between the implant stability quotient (*ISQ*) and bone implant contact (BIC) nor between *ISQ* and cortical thickness has been evidenced so far [22]. Finite element numerical simulation tools showed that the orientation and fixation of the transducer have an important effect on *ISQ* values [20] obtained with the older Osstell version with an L-shaped, wired transducer.

An alternative method has been developed by our group and consists in using a quantitative ultrasound (QUS) method [23] to investigate the properties of the bone-implant interface. The principle of the measurement relies on the dependence of ultrasonic propagation within the implant on the boundary conditions given by the biomechanical properties of the bone-implant interface [24]. An *in vitro* preliminary study was carried out using a prototype titanium cylinder shaped implants inserted in bone tissue, showing the sensitivity of the ultrasound response of the implant to the quantity of bone in contact with the implant [24]. The principle of the measurement was then validated experimentally by showing the sensitivity of the echographic response of a planar bone-implant interface to healing time using coin-shaped implant models [25]. Significant variations of the ultrasonic response of dental implants embedded in a bone substitute biomaterial (a tricalcium silicate-based cement) was shown to occur when the implants are subjected to fatigue loading [26]. More recently, another *in vitro* study proved the potentiality of QUS methods to assess dental implant primary stability [27]. A preclinical validation of the device in rabbits was carried out [25, 28] and showed that i) the measurement was sensitive to healing time and ii) a significant correlation of the measurement with the bone-implant contact (BIC) ratio measured with histology. Moreover, finite difference [29] and finite element numerical simulations [30, 31] were carried out to understand the interaction between an ultrasonic wave and the bone-implant system, leading to a better performance of the device.

The comparison of the experimental results obtained in a controlled configuration using the QUS and RFA techniques would be of interest in order to better assess the performance of the different approaches. The aim of the present study is to compare these the RFA and QUS methods which have been evoked to assess dental implant stability. To do so, our strategy consists in using dental implants inserted in bone mimicking phantoms made of polyurethane because it allows to work under standardized and reproducible conditions. Different parameters related to the implant stability (such as the density of the bone phantom, the thickness of cortical bone, the insertion depth and the drill diameter) were investigated and the related variations of different quantities such as i) the RFA response and ii) the ultrasonic response were investigated.

## II) Materials and Methods

**II)1 Bone mimicking phantoms and dental implant**

The bone mimicking phantoms used in the present study were made of rigid polyurethane foam (Orthobones®, 3B Scientific, Hamburg, Germany) with different values of bone density and of cortical thickness (1 and 2 mm). Cortical bone was modeled by the material type #40 PCF with a mass density equal to 0.55 g/cm³. Three types of trabecular mimicking phantoms were considered (#10, #20, #30 PCF) with mass density values equal to 0.16 g/cm³, 0.32 g/cm³ and 0.48 g/cm³, respectively, according to the manufacturer.

All dental implants used in this study were commercial implants manufactured by Zimmer Biomet (Winterthur, Switzerland) under the reference TSVT4B10. The geometrical characteristics are a length of 10 mm and an external diameter of 4.1 mm.

A conical cavity was then created in the block bone test, using color-coded, 10-mm-length surgical drills preconized by Zimmer Biomet and used as a reference protocol by dentists. Different values were considered for the final drill diameter, as described in details in subsection II)3.

**II)2 Measurement methods of dental implant stability**

Two different and complementary methods were used in order to measure the implant stability of each implant in each configuration, which are described in subsection II)3.

*a) Resonance frequency analysis*

For each implant and each configuration, the RFA response of the implant was measured in *ISQ* units (on a scale from 1 to 100) using the Osstell device (Osstell, Göteborg, Sweden). Figure 1a shows the configuration of the measurements which were realized using a smart peg placed in the implant, as recommended by the manufacturer. Each measurement was performed 5 times (in order to assess the reproducibility of the measurements) in two perpendicular directions denoted 0° and 90° relatively to an arbitrary axis chosen for each block. The average and standard deviation values $ISQ_m$ and $ISQ_{std}$ of the ten different *ISQ* values were determined for each implant and each configuration.

*b) Quantitative ultrasound device*

Figure 1b shows the dedicated QUS device, which consists in a 5 mm diameter planar ultrasonic contact transducer (Sonaxis, Besançon, France) generating a 10 MHz broadband (the frequency bandwidth is approximately equal to 6–14 MHz) ultrasonic pulse propagating perpendicularly to its active surface (monoelement transducer used in echographic mode). The probe was attached rigidly to a titanium alloy dental healing abutment which can be screwed into the implant so that the measurements are not influenced by positioning problems of the probe relatively to the healing abutment, as it was the case in previous studies [26]. The QUS device was screwed into the implant in order to realize each measurement for each implant and each configuration, as shown in Fig. 1b. The ultrasonic probe was linked to a pulser-receiver via a standard coaxial cable. A transient recorder was used to record the radiofrequency (*rf*) signal with a sampling frequency equal to 100 MHz.

For each measurement, the transducer was screwed in the implant with a torque of 0.035N.m, which is around 10 times lower than torque values recommended by implant manufacturers [32] and the echographic measurement was made instantaneously. The transducer was then unscrewed and the same measurement was carried ten times in order to assess the reproducibility of the measurements.

The same signal processing technique as the one used in [26] was used to derive a quantitative ultrasonic indicator *UI* which had been shown to be related with dental implant stability. The indicator *UI* was devised to quantitatively estimate the average amplitude of the signal between 10 and 120 µs. Therefore, the envelop *S(t)* of the *rf* signal *s(t)* was determined and the indicator *UI* is given by:

$$UI = \sum_{i=1000}^{12000} S(iT_0), \quad (1)$$

where $T_0$=0.01 µs corresponds to the sampling period. The time window chosen to determine the indicator *UI* was chosen as follows. The upper bound of the time window equal to 120 µs was chosen, which corresponds to a compromise between a sufficient duration to obtain pertinent information and the requirement of a sufficient signal to noise ratio for all *rf* signals. Moreover, the time window used to compute the indicator *UI* starts at a time of 10 µs because the amplitude of the envelop of the *rf* signals before 10 µs is approximately constant due to a saturation of the amplitude.

The average and standard deviation values $UI_m$ and $UI_{std}$ of the indicator *UI* obtained for the ten measurements were determined for each implant and each configuration.

**II)3 Experimental protocol**

The effect of varying the different parameters described below on the variation of the *ISQ* and the indicator *UI* was investigated in this study, as summarized in Figure 2. For each configuration described below, the values of the averaged and standard deviation values $ISQ_m$ and $ISQ_{std}$ (respectively $UI_m$ and $UI_{std}$) of the *ISQ* (respectively the ultrasonic indicator *UI*) were measured.

   a) *Variation of trabecular bone density*

The effect of modifying trabecular bone properties was first investigated by comparing the results obtained with the two methods described in subsection II)2 for implants fully inserted in different bone mimicking phantoms having the same cortical thickness equal to 2 mm, which corresponds approximately to the clinical situation [33]. Three different materials were used to mimic trabecular bone (#10, #20 and #30 PCF). Three implants were inserted with a final drill of 3.4 mm of diameter in each type of bone mimicking phantom, resulting in a total of 9 configurations.

   b) *Variation of cortical bone thickness*

The effect of changing the cortical bone thickness was then investigated because cortical thickness is known to be an important determinant of implant stability [11]. To do so, two bone mimicking phantoms with two different cortical thicknesses (1 and 2 mm) and the same material to mimic trabecular bone (#10 PCF) were considered. Again, three implants were fully inserted with a final drill of 3.4 mm of diameter in each type of bone mimicking phantom, resulting in a total of 6 configurations.

   c) *Variation of the diameter of the final drill*

The effect of the diameter of the final drill used to machine the cavity in two given bone mimicking phantom was investigated. The two phantoms used had the same cortical thickness (1 mm) and trabecular bone was mimicked by #10 and #30 PCF. Ten values of diameter of the final drill comprised between 2.7 and 3.6 mm with a step of 0.1 mm were considered. Three implants were considered for each diameter tested, leading to a total number equal to 30 configurations.

### d) Variation of the implant insertion depth

Eventually, the effect of the implant insertion depth was investigated because it corresponds to a simple way of imposing a variation of dental implant stability, similarly as what has been done in [28]. To do so, three cavities were realized in a given bone mimicking phantom (cortical thickness of 1 mm and #30 PCF for trabecular bone) with a final drill diameter of 3.4 mm. Then, an implant was screwed in each cavity so that half of the implant was inserted into the cavity. The average and standard deviation values of the *ISQ* and of the ultrasonic indicator *UI* were measured. The implant was then screwed by $\pi$ rad in order to increase the implant surface area in contact with bone mimicking phantom. The two measurements described above (*ISQ* and *UI*) were realized again. The procedure was repeated until the implant was completely inserted in bone mimicking phantom, this screwing level is denoted "0". Then, the implant was then unscrewed by $\pi$ rad in order to reduce the surface area of the implant in contact with bone block. The same two measurements (*ISQ* and *UI*) were again carried out. The procedure was reproduced until the implant was completely detached from the test block.

## II)4 Statistical analyses

A one-way analysis of variance (ANOVA) and Tukey-Kramer tests were performed to evaluate the significance of variations of the ultrasonic indicator *UI* and the *ISQ* as a function of the different configurations. Statistical differences were defined at a 95% confidence level.

# III) Results

## III)1 Effect of bone density

Figure 3 shows the results obtained for the nine configurations corresponding to three models of trabecular bone quality (#10, #20 and #30 PCF) with the different methods of implant stability assessment (*ISQ* and *UI*). As shown in Fig. 3, the values of $ISQ_m$ increase when the density of trabecular bone increases, while the opposite behaviour is obtained for the ultrasonic indicator $UI_m$. The error bars represent the reproducibility of the ultrasound and RFA measurements for each configuration. The average value of $UI_{SD}$ (respectively $ISQ_{SD}$) obtained for the ultrasonic measurement (respectively RFA measurements) is equal to 0.07 unit (respectively 4 ISQ units).

ANOVA shows a significant effect of bone density on the values of *UI* (p-value < $10^{-8}$) and of *ISQ* (p-value = 0.04). Moreover, Tukey-Kramer tests show that i) the results obtained for *UI* in #10 PCF (respectively #20 PCF) are significantly different than the results obtained in #20 PCF, p-value < $10^{-4}$ (respectively #30 PCF, p-value < $10^{-7}$). The results obtained with the RFA measurements are significantly different when comparing implants inserted i) in #10 PCF and #20 PCF (p-value < $3.9*10^{-8}$) and ii) in #10 PCF and #30 PCF (p-value < $2.6*10^{-8}$). However, the results obtained with RFA measurements in #20 PCF and #30 PCF are not significantly different (p-value = 0.93).

## III)2 Effect of cortical thickness

Figure 4 shows the variation of the *ISQ* and of the *UI* for different implants inserted in test blocks with 1 mm and 2 mm of cortical thickness. Trabecular bone density is equal to #10 PCF.

ANOVA shows a significant effect of cortical thickness on the values of the *UI* (p-value < $10^{-9}$) and on the values of *ISQ* (p-value = 0.04), although the p-value obtained for *ISQ* is relatively high.

**III) 3 Effect of the final drill diameter**

Figure 5 shows the variation of the average and standard deviation values of the ISQ and of the *UI* as a function of the final drill diameter for implants inserted in the same test block with a cortical thickness of 1 mm and a density of trabecular density equal to #10 PCF. The results show that the values of the *UI* (respectively *ISQ*) increase (respectively decrease) as a function of the final diameter drill. In order to clarify the figure, we have chosen to present only one curve (instead of three) because the other results are qualitatively similar to those shown in Fig. 5.

ANOVA shows a significant effect of final drilling diameter on the values of the *UI* (p < $10^{-12}$) and on the values of *ISQ* (p < $10^{-3}$). The stars in Fig. 5 indicate the configurations for which the results obtained for the *UI* and for the *ISQ* are statistically similar, which was obtained using a Tukey-Kramer test. The results obtained with Tukey-Kramer test indicate that the *UI* obtained for all final drill diameters are significantly different to the other ones, except for the results obtained for the final drill diameters equal to 2.8 and 2.9. However, the results obtained for *ISQ* with all drill diameters lower than 3.3 (respectively higher that 3.2) are statistically similar.

**III) 4 Effect of the insertion depth**

Figure 6 shows the average and standard deviations of the *ISQ* and of the *UI* for the same implant with different values of the penetration depth. The arrows in Fig. 5 indicate the evolution of the results when the implant is screwed (from a bone level of -5π rad to 0 rad) and then unscrewed (from a bone level of 0 rad to -4π rad). The *UI* value is shown to decrease during the insertion of the dental implant in the bone block while the *UI* value increases when the implant in unscrewed from the test block. The *ISQ* values increase during the insertion stages and then decrease when the implant is unscrewed. In order to clarify the figure, we have chosen to present only one curve (instead of three) because the results obtained using the other test block are qualitatively similar to those shown in Fig. 6.

## IV) Discussion

The originality of the present study was to compare two different techniques (RFA and ultrasound measurements) in order to investigate the primary stability of dental implants inserted in artificial bone blocks (Orthobones®). Different stability conditions were considered by varying different parameters such as the type of bone block (trabecular bone density, cortical thickness), the final diameter drill and the insertion depth, in order to simulate different situations mimicking variations of dental implant primary stability.

**IV) 1 Physical interpretation and comparison with the literature**

The values of the *ISQ* are shown to increase when i) trabecular density increases (see Fig. 3), ii) cortical thickness increases (see Fig. 4), and iii) the insertion depth increases (see Fig. 6). These results are in agreement with previous studies [11, 14, 17, 34] and can be explained by the higher rigidity of the block-implant system induced by an increase of trabecular density, of cortical thickness and of implant insertion depth.

The values of the *UI* are shown to decrease when i) trabecular density increases (see Fig. 3), ii) cortical thickness increases (see Fig. 4), iii) the final diameter drill decreases (see Fig. 5) and

iv) the insertion depth increases (see Fig. 6). These results are in agreement with previous papers by our group that showed that i) the *UI* also increases when a dental implant is unscrewed in a bone sample in *ex vivo* [27] as well as *in silico* [31] studies, ii) the *UI* decreases when the Bone-Implant contact ratio increases for *in vivo* measurements [28] and iii) the *UI* decreases when trabecular bone quality increases and when cortical thickness increases for *in silico* measurements [29-31]. The aforementioned results can be explained by the fact that the *UI* is related to the amplitude of the echographic response of the implant, which depends on the boundary conditions applied to the implant external surface. The implant acts as a wave guide in which the acoustical energy is trapped [29-31]. When bone quantity and quality increase around the implant surface, the gap of mechanical properties between the implant and its environment decreases, which induces an increase of the transmission coefficient. Therefore, the ultrasound energy decreases faster since ultrasound may propagate in the surrounding medium, resulting in a faster decrease of the implant echographic response. This physical interpretation may qualitatively explain the results corresponding to the variation of the *UI* obtained herein.

Figure 6 shows that the values of the *UI* (respectively *ISQ*) are lower (respectively higher) for the same insertion value during the screwing phase compared to the unscrewing phase. These results may be explained by damage and wear phenomena occurring during the implant insertion, which lead to a higher bone-implant contact ratio during the screwing phase compared to the unscrewing phase, thus explaining the lower (respectively higher) values of *UI* (respectively *ISQ*) during the insertion phase compared to the unscrewing phase.

The results obtained in the present study depend on the displacement of the implant surface generated by the ultrasound transducer, which is difficult to measure using an experimental approach because accessing the surface of an implant embedded in a bone mimicking phantom remains a difficult task. However, the analysis of the interaction between an ultrasonic wave and a dental implant in a situation similar to the one described herein was carried out in previous in silico studies by our group [29,30,31]. Briefly, the importance of the lateral wave was evidenced in preliminary studies considering simple cylindrical shaped implants [29,30]. In a more recent study considering the geometry of an actual dental implant [31], the sensitivity of the ultrasonic indicator to changes occurring around 30 µm around the implant surface was evidenced.

**IV) 2 Comparison between the RFA and QUS methods**

The results obtained in this study allow to compare quantitatively the ultrasound and RFA approaches to assess dental implant primary stability in controlled configurations.

a) Drill diameter

As shown in Fig. 5, the values obtained for the *UI* are significantly different when all final drill diameters vary between 2.7 mm and 3.6 mm (except for 2.8 and 2.9 mm) whereas the values obtained for the *ISQ* are similar when the final drill diameter varies between 2.7 mm and 3.2 and when it varies between 3.3 and 3.6. This results show that the ultrasound device can discriminate implants inserted in test blocks with a wider range of final diameter size compared to the RFA measurements.

b) Sensitivity analysis

The results can be analyzed in order to determine the sensitivity of QUS and RFA methods to the different parameters related to the implant stability. To do so, a simple two step method is described in what follows (refer to [27] for further details on this method) and constitutes a simple approach in order to estimate the sensitivity of both methods to a variation of a given parameter *X* related to the implant primary stability. Note that *X* may correspond to trabecular bone density, to cortical thickness or to the implant insertion depth. The sensitivity of the *ISQ* to variations of the final drill diameters was not investigated herein because the results obtained with the RFA techniques are weakly sensitive to the drill diameter (see subsection a) above).

The first step is to perform a linear regression analysis of the average value of the *ISQ* (respectively of the *UI*) as a function of the parameter *X* by analyzing the results shown in Fig. 3-6, which leads to the following approximated relations:

$$\widetilde{ISQ} = a_{ISQ}X + b_{ISQ} \qquad (2)$$

$$\widetilde{UI} = a_{UI}X + b_{UI}, \qquad (3)$$

where *X* is the investigated parameter, $\widetilde{ISQ}$ and $\widetilde{UI}$ are the approximated values of the *ISQ* and the *UI* respectively and $a_{ISQ}$ and $b_{ISQ}$ (respectively $a_{UI}$ and $b_{UI}$) are the coefficients found by applying a linear regression analysis to the variation of the *ISQ* (respectively the UI) as a function of *X*.

The second step of the method consists in using the averaged reproducibility error corresponding to a given configuration in combination with the linear regression analyses corresponding to Eqs. 2 and 3, in order to assess the error realized on the estimation of the parameter *X*, noted in what follows $X_{ISQ}$ (respectively $X_{UI}$) for the error realized using the RFA (respectively ultrasound) technique:

$$X_{ISQ} = ISQ_{std} / a_{ISQ} \qquad (4)$$

$$X_{UI} = UI_{std} / a_{UI} \qquad (5)$$

As expected, the error on the estimation of the parameter *X* increases when the reproducibility error (given by $ISQ_{std}$ and $UI_{std}$) increases and when the sensitivity of the method (given by $a_{ISQ}$ and $a_{UI}$) decreases.

Table 1 shows the values of the error made on the determination of the trabecular density, of the cortical thickness and of the insertion depth obtained by applying the aforementioned procedure on the data shown in Fig. 3, 4 and 6 respectively. As shown in Table 1, the error made on the estimation of the trabecular density using the ultrasound method is more than four times lower compared to that using the RFA analysis, which was obtained by analyzing the results obtained in Fig. 3. Note that the values of *ISQ* obtained for #20 and #30 PCF test blocks were not significantly different, while the results obtained for the *UI* for #20 and #30 PCF test blocks were significantly different. Therefore, the quantitative ultrasound method is more sensitive than the RFA method to retrieve the trabecular density.

Table 1 also shows the results corresponding to the error realized on the cortical thickness estimation using the data shown in Fig. 4. The results indicates that the error realized on the estimation of the cortical thickness using the ultrasound method is approximately eight times lower compared to that using the RFA analysis. Note that the values of *ISQ* obtained with cortical thicknesses equal to 1 and 2 mm were not significantly different while the results obtained for the *UI* obtained with cortical thicknesses equal to 1 and 2 mm were significantly different. Therefore, the quantitative ultrasound method is more sensitive the RFA analysis to retrieve the cortical thickness.

Table 1 shows the results corresponding to the error realized on the insertion level estimation using the data shown in Fig. 5. The results indicates that the error realized on the estimation of the insertion level using the ultrasound method is approximately four times lower compared to that using the RFA analysis. Therefore, the quantitative ultrasound method is more sensitive the RFA analysis to retrieve the cortical thickness.

As shown in Table 1, the performances of the quantitative ultrasound device to retrieve dental implant primary stability are better than the performances of RFA technique in the simple configurations considered herein. These results can be explained by the fact the ultrasound device is sensitive to bone properties at the intimate contact with the implant, which is precisely the important parameter determining the implant stability [1, 35, 36]. More specifically, it has been shown in a previous *in silico* study [30] that the ultrasound technique is sensitive to the bone properties at a distance lower than around 30 µm. In the case of the RFA technique, the frequency response of the implant is sensitive to the biomechanical properties of the bone-implant interface as well as to the surrounding medium since the entire test block is likely to be excited by the RFA device. Although the amplitude of the displacement field induced by the RFA method is small compared to the implant size, the excitation mode corresponds to the first resonance of the bending mode, which indicates that a vibration is induced by the transducer. This vibration is applied to the bone-implant system as a whole (and not only to the implant interface) through the implant surface. However, in the case of the QUS device, the excitation induced by the transducer differs from the RFA device because propagation phenomena should be considered [31], instead of resonance effects. The ultrasound wave propagating within the implant [31] is therefore sensitive to its interface only, which has been quantified in [31].

Moreover, the results obtained with the RFA device depend on the direction of solicitation, which is not the case of the ultrasound device. In the case of the RFA method, the dependence of the results on the orientation of the transducer has been explained by the anisotropic nature of bone tissue but it remains difficult to extract useful information from such variation [37, 38].

**IV) 3 Limitations of the study**

The main limitation of this study is that bone mimicking phantoms were considered, which allows to work under standardized and reproducible conditions. Note that both techniques had been validated previously independently *in vivo* [28, 34]. However, the artificial bone blocks are adapted to mimic the healthy jaw bone because the mean bone mineral density was 0.31 g/cm³ for the posterior maxilla and 0.55 g/cm³ for the anterior maxilla [39]. For healthy jaw bone, the mean cortical thickness for the mandible is 2.22±0.47 mm and the mean cortical bone thickness for the maxilla is 1.49±0.34 mm [33]. Further *in vivo* validations should be carried out clinical studies should also be made.

Another limitation consists in the fact that only one implant type was used in this study, because the goal was to investigate the effect of variations of the implant stability on the *UI* and on the *ISQ*. It would be of interest to carry out the same study with other implant types. However, previous studies [26, 28] have shown the validation of the ultrasound device using other implant types.

The present study does not address the important clinical question corresponding to the determination of the extent to which the RFA and QUS methods can actually accurately measure an important information from a clinical point of view. The question of the relationship between the accuracy of each method versus the precision needed in the context of implant stability remains open and is not in the scope of the present study. Further clinical studies are needed to determine whether either method (RFA or QUS) can accurately measure a clinically relevant information.

## Conclusion

This study allows to compare the results obtained with two different approaches aiming at estimating primary dental implant stability, which are realized with the same implants under various configurations. Namely, we considered a variation of the following parameters: i) trabecular bone density, ii) cortical thickness, final drill diameter and iv) penetration depth. All results are consistent and can be explained by physical analyses of the biomechanical phenomena occurring around the implant. Moreover, we found that ultrasound technique provides a better estimation of different parameters related to the implant stability compared to RFA techniques. Therefore, the present study paves the way for the development of an ultrasonic device to estimate dental implant stability that could be used in the clinic provided further *in vivo* investigations.

# Acknowledgements

This work was supported by the French National Research Agency (ANR) through the PRTS program (project OsseoWave n°ANR-13-PRTS-0015-02). This work has received funding from the European Research Council (ERC) under the European Union's Horizon 2020 research and innovation program (grant agreement No 682001, project ERC Consolidator Grant 2015 BoneImplant).

# Tables

| Estimated parameter X | Indicator | a | b | Indicator standard deviation | Error on the estimation |
|---|---|---|---|---|---|
| Trabecular density | *ISQ* | 0.55 | 56.50 | 1.5 | 2.73 |
| | *UI* | -0.13 | 4.73 | 0.08 | 0.6 |
| Cortical thickness | *ISQ* | 3.6 | 54.4 | 1.1 | 0.31 |
| | *UI* | 2.70 | 8.70 | 0.1 | 0.04 |
| Insertion depth | *ISQ* | 7 | 73.9 | 1 | 0.16 |
| | *UI* | -2.5 | 0.1 | 0.1 | 0.04 |

Table 1. Results obtained for the linear regression analysis (*a* and *b*), the average standard deviation of each indicator and for the error realized on the estimation of each parameter *X*.

Figure Legends

Figure 1: Description of the experimental protocol. Measurement realized using a) the resonance frequency analysis device, which leads for a given implant to an average and standard deviation value of the score ($ISQ_m$ and $ISQ_{std}$) and b) the quantitative ultrasound device, which leads for a given implant to an average and standard deviation value of the score ($UI_m$ and $UI_{std}$).

Figure 2: Summary of the experimental protocol describing the parameters investigated in this study. For each series of experiments, the parameters which vary are indicated in bold.

Figure 3: Variation of the values of the *ISQ* and of the *UI* for implants inserted in test blocks with different values of the trabecular density (#10, #20 and #30 PCF). Three implants are considered per test block.

Figure 4. Variation of the values of the *ISQ* and of the UI for 3 implants inserted in a test block with a cortical thickness of 1 mm and for 3 implants inserted in a test block with a cortical thickness of 2 mm.

Figure 5. Variation of the values of the *ISQ* and of the UI for implants inserted in cavities obtained with different values of the final drill diameter. The stars indicate the results that are statistically similar. The errorbars correspond to reproducibility of each measurement.

Figure 6. Variation of the ultrasonic indicator and *ISQ* values for different screwing levels corresponding to the same implant inserted in a test block with 1 mm of cortical thickness and a trabecular bone density of #30 PCF.

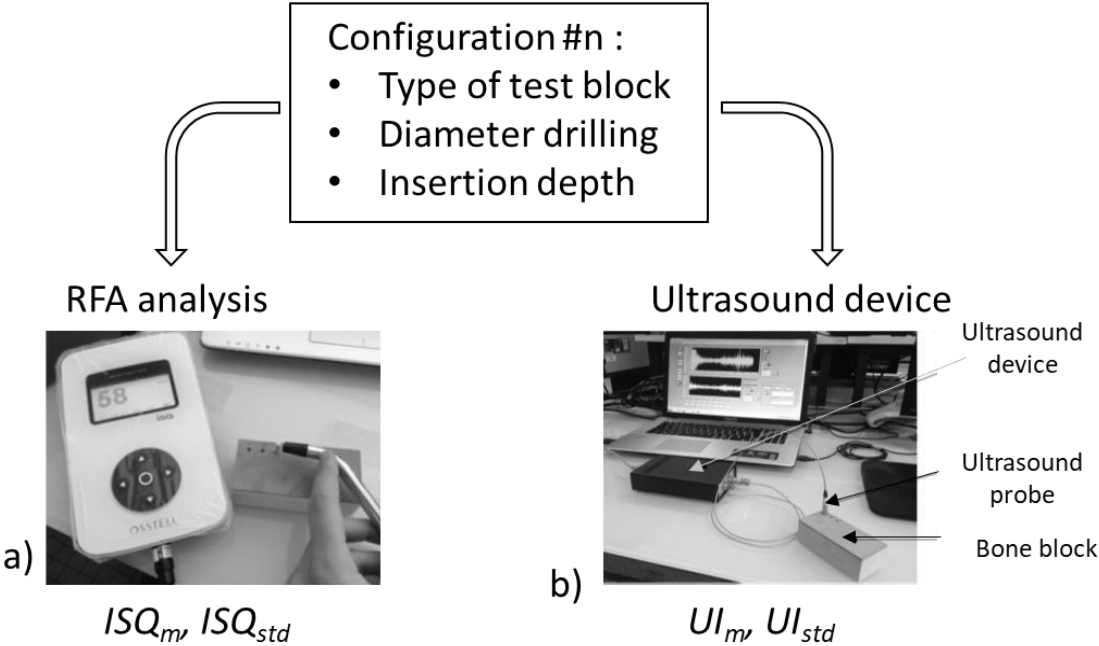

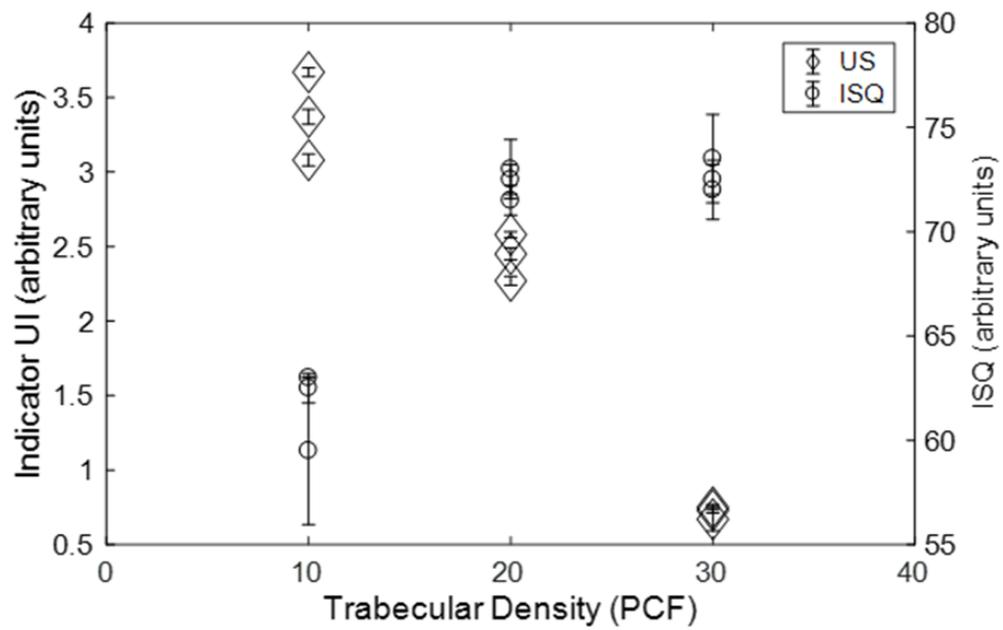
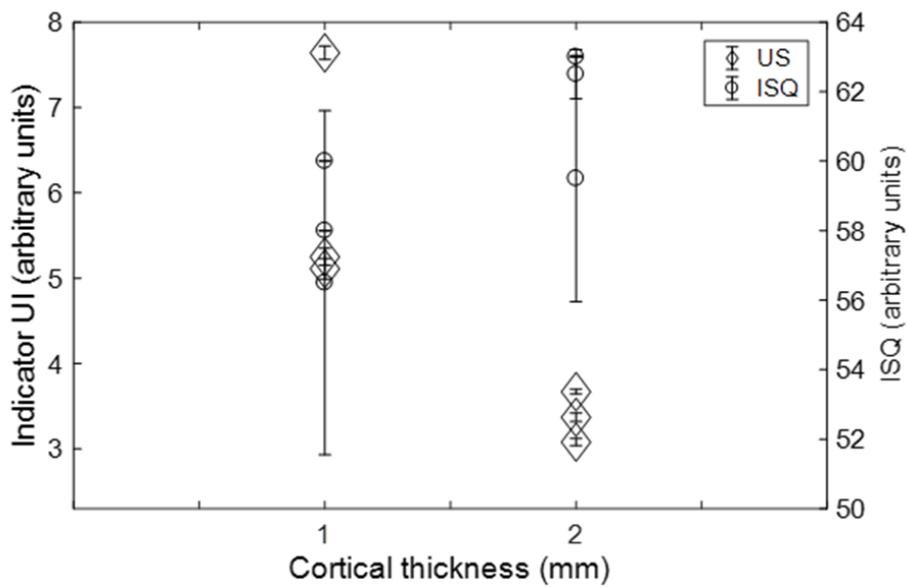

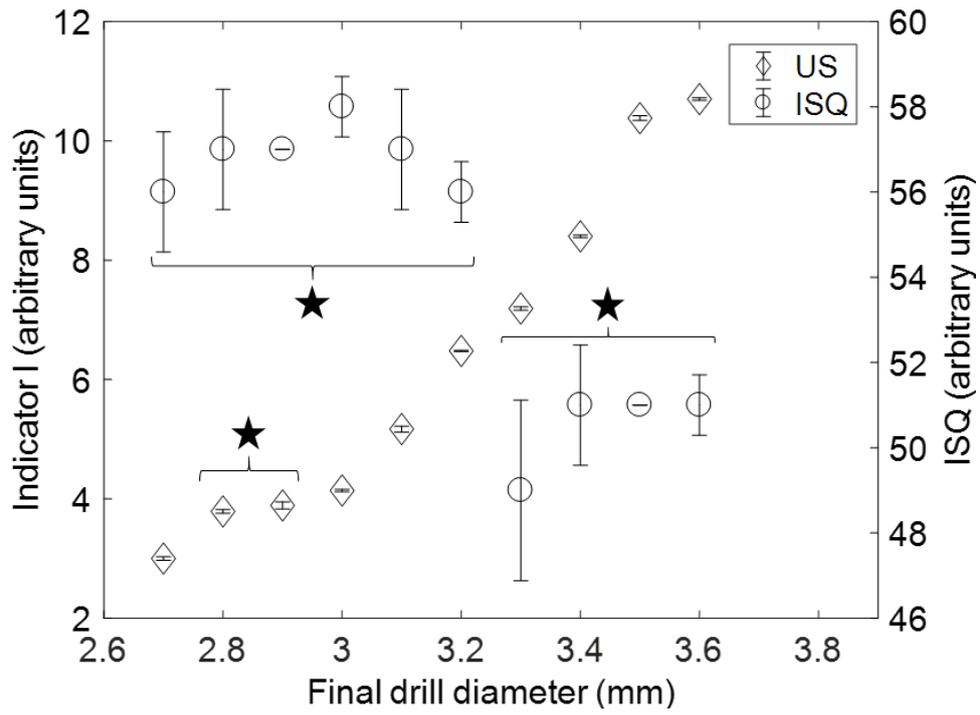

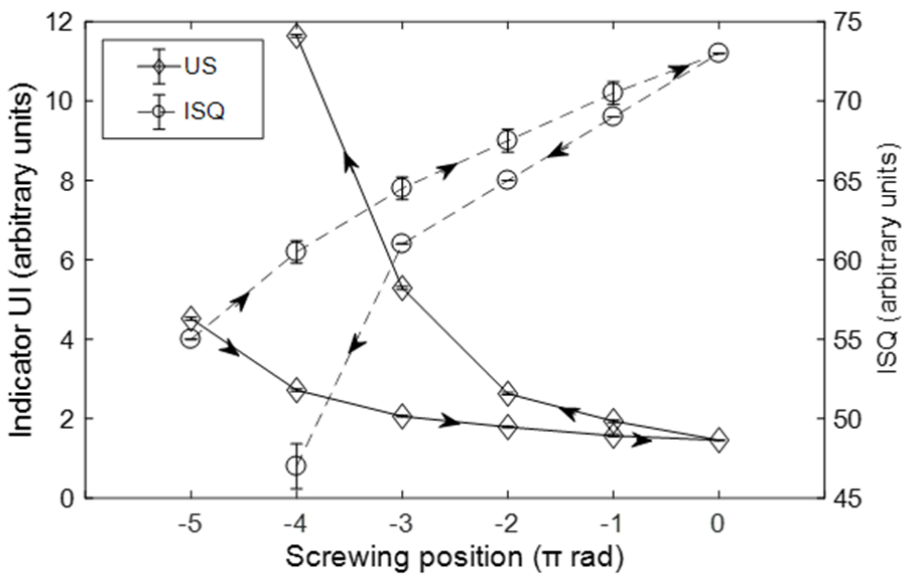